\documentclass[twocolumn,showpacs,preprintnumbers,amsmath,amssymb]{revtex4-1}

\usepackage{graphicx}
\usepackage{dcolumn}
\usepackage{bm}
\usepackage{soul, xcolor}
\usepackage{color, wasysym}
\usepackage{textcomp}
\usepackage{amsmath, amssymb, amsthm, latexsym, epsfig}

\DeclareMathOperator{\D}{d\!}
\DeclareMathOperator{\E}{e} 
\DeclareMathOperator{\I}{i}
   \DeclareMathOperator{\RE}{\mathfrak{Re}}

\begin{document}

\newtheorem{theorem}{Theorem}
\newtheorem{definition}{Definition}
\newtheorem{lemma}{Lemma}
\newtheorem{proposition}{Proposition}
\newtheorem{remark}{Remark}
\newtheorem{con}{Conjecture}
\newtheorem{example}{Example}


\title{Non-Debye relaxations: The characteristic exponent in the excess wings model}

\author{K. G\'{o}rska}
\email{katarzyna.gorska@ifj.edu.pl}
\author{A. Horzela}
\email{andrzej.horzela@ifj.edu.pl}
\affiliation{Institute of Nuclear Physics, Polish Academy of Sciences, \\ ul. Radzikowskiego 152, PL-31342 Krak\'{o}w, Poland}

\author{T. K. Pog\'{a}ny}
\affiliation{Faculty of Maritime Studies, University of Rijeka, 51000 Rijeka, Croatia \\
Institute of Applied Mathematics, \'Obuda University, H-1034 Budapest, Hungary}
\email{poganj@pfri.hr}


\begin{abstract}
The characteristic (Laplace or L\'evy) exponents uniquely characterize infinitely divisible probability distributions. Although of purely mathematical origin they appear to be uniquely associated with the memory functions present in evolution equations which govern the course of such physical phenomena like non-Debye relaxations or anomalous diffusion.  Commonly accepted procedure to mimic  memory effects is to make basic equations time smeared, i.e., nonlocal in time. This is modeled either through the convolution of memory functions with those describing relaxation/diffusion or, alternatively, through the time smearing of time derivatives. Intuitive expectations say that such introduced time smearings should be physically equivalent. This leads to the conclusion that both kinds of so far introduced memory functions form a ``twin'' structure familiar to mathematicians for a long time and known as the Sonine pair. As an illustration of the proposed scheme we consider the excess wings model of non-Debye relaxations, determine  its evolution equations and discuss properties of the solutions.
\end{abstract}

                             

\maketitle

\section{Introduction}\label{sec0}

Typical example of dielectric relaxation is provided by a dipolar system which approaches  the equilibrium being earlier driven out of it by a step or alternating external electric field. The phenomenon is usually described in terms of the relaxation function $n(t)$ which counts dipoles surviving depolarization during the time $(0,t)\subset(0,\infty)$ and, if normalized, evolves form $n(0+) = 1$ to $n(\infty) = 0$. The  function $n(t)$ comes out as the solution of macroscopic differential equation
\begin{equation}\label{6/08-2}
\dot{n}(t) = - r(t, \tau) n(t). 
\end{equation}
The non-negative quantity $r(t, \tau)$ is the transition rate of the system and besides of the time depends on properties characterizing the medium  among which a material constant called the relaxation, or characteristic, time $\tau$ is the most important. Solution to Eq. \eqref{6/08-2} is easily got as $n(t) = \exp[-\int_{0}^{\infty}r(\xi, \tau)\D \xi]$ but it remains of very limited physical utility because the knowledge of $r(t, \tau)$, especially for short and long times $t$, is far insufficient except of the Debye case for which $r(t, \tau) = {\tau}^{-1}={\rm const}$. Data provided by the broadband dielectric spectroscopy (encoded in the so--called spectral functions) extrapolated to the full frequency range and next transformed to the time domain, do not help very much - using them to calculate the ratio $\dot{n}(t)/n(t)$  usually leads to cumbersome formulae \cite{Gloeckle,RHilfer02,RHilfer02a,KGorska18}, in addition singular at the origin, which makes their experimental verification rather impossible for the time laps close to the origin. These difficulties have prompted efforts to look for mesoscopic description of relaxation phenomena based on dynamical rules being non-local in time and leading to the evolution equations which from the very beginning take into account the memory effects. The simplest way to mimic the memory is to introduce the time smearing which may proceed two-fold: either one smears the left hand side of Eq. \eqref{6/08-2}, i.e., the time derivative in $\dot{n}(t) = - r(t, \tau) n(t)$ or rewrites Eq. \eqref{6/08-2} in the integral form ${n}(t) =1 - \int_{0}^{t}r(\xi, \tau) n(\xi)\D \xi$ and uses the smearing $r(\xi, \tau)\to r(t-\xi, \tau)$. Keeping the relaxation time $\tau$ explicitly separated out from the other material depending parameter's vector ${p}$, say, this leads to 
\begin{equation}\label{19/02-2}
      \int_{0}^{t} k(t - \xi)~ \dot n(\xi)\, \D\xi = - B(\tau; {p})~ n(t)\,,
   \end{equation}
where $B(\tau, {p})$ denotes the universal, time independent, transition rate and $k(t)$ stands for the memory kernel responsible for smearing the time derivative. In turn, for  $r(t - \xi, \tau) = B(\tau; {p}) M(t - \xi)$, we get 
\begin{equation}\label{19/02-1}
      n(t) = 1 - B(\tau, {p}) \int_{0}^{t} M(t - \xi)~ n(\xi)\, \D\xi,
   \end{equation}
with $M(t - \xi)$ being another memory kernel, \textit{a priori} not connected to $k(t-\xi)$. Mathematically Eqs. \eqref{19/02-2} and \eqref{19/02-1} are both the Volterra type equations \cite{GGripenberg} which utility goes beyond more popular fractional differential equations introduced in the framework of fractional calculus approach to the relaxation phenomena \cite{KGorska20}. If we require physically justified equivalence of Eqs. \eqref{19/02-2} and \eqref{19/02-1} then the memory kernels become mutually related and form a coupled (Sonine) pair which appearance and properties we shall discuss a bit later.

The integro-differential equation \eqref{19/02-2} is mathematically very well understood \cite{ANKochubei11}. In fact it is the equation which for special choices of $k(t-\xi)$ reduces to equations with fractional derivatives (more precisely, various types of them) for a long time proposed to investigate the  relaxation phenomena. Simultaneously, both Eqs. \eqref{19/02-2} and \eqref{19/02-1} have the form of kinetic equations which are the starting point to describe relaxation in the subordination framework \cite{AStanislavsky15}, \cite{AStanislavsky17}, \cite[Chs. 4.1, 4.3]{AStanislavsky19_1} developed as a general scheme within the stochastic processes approach to the relaxation phenomena and anomalous diffusion.

The cornerstone of the stochastic processes based approach to relaxation (as well as to anomalous diffusion if one adopts a suitably reinterpreted language) is the assumption that the transition rate $r(t,\tau)$ introduced in Eq. \eqref{6/08-2} takes on the meaning of a non-negative stochastic quantity parametrized by the randomized characteristic time $\tau$. The latter does not denote any longer one among material properties of the relaxing medium and becomes physically meaningful variable which shape of postulated randomization strongly influences, or even determines, modeling the relaxation. The choice of stochastic processes proposed to investigate relaxation phenomena is domi-\\nated by choosing those which are non-negative, non-decreasing and have distributions which are infinitely divisible. The last means that relavant distributions functions $f(\widetilde{\beta})$ are representable as $N\to \infty$ limit of distributions obeyed by random variables $\widetilde{\beta}^{(N)} = \frac{1}{N}\sum_{i=1}^{N}\lambda_{i}$ where all $\lambda_{i}$ are independent identically distributed random variables \cite{L1}.  Randomization of the characteristic time $\tau$  (in the Debye systems assumed to be the same and fixed for all dipoles forming the system) means that we are going to change description of the system - instead of looking for deterministic evolution in the time $t$ measured by a laboratory clock  we search for stochastic evolution  in terms of the ``internal'' time $\tau(t)$ whose dependence on $t$ is hidden in some probability distribution $f(\tau,t)$.

Any non-negative stochastic process whose distribution is infinitely divisible, herewith denoted as $U(\tau)$,  satisfies the relation  
\begin{equation}\label{27/07-1}
\left\langle\exp{(-sU(\tau))}\right\rangle=\exp{(-\tau\widehat{\Psi}(s))},
\end{equation}
where $\widehat{\Psi}(s)$ bears the name of characteristic (either Laplace or L\'{e}vy) exponent and is uniquely given by the L\'evy--Khintchine formula \cite[Eq. (1.3)]{Schilling}
\begin{equation}\label{27/07-2}
\widehat{\Psi}(s)=\lambda s + \int_{0}^{\infty}(1 - \E^{-sx}) \mu(\D x)
\end{equation}
where $\mu(\D x)$ (subject to some additional conditions) is called the L\'evy measure while $\lambda$ is named the drift parameter. For $s>0$ the relation \eqref{27/07-2} places all functions $\widehat{\Psi}(s)$ in the class of Bernstein functions (BFs), \textit{i.e.} non-negative functions on $\mathbb{R}_{+}$, differentiable infinitely many times and satisfying for $s>0$ and $n\in\mathbb{N}_{0}$ the conditions $(-1)^{n}f^{(n+1)}(s)\ge 0$  everywhere in their domain  \cite{L2}. We remark that the BFs are close relatives to the completely monotone functions (CMFs), also being non-negative on $\mathbb{R}_{+}$, differentiable there infinitely many times and satisfying $(-1)^{n}f^{(n)}(s)\ge 0, s>0, n\in\mathbb{N}_{0}$. To make these notions more intuitive one may understand BFs as ``maximally regularly'' increasing positive functions  while CMFs as ``maximally regularly'' decreasing ones \cite{L3}.
The deep mutual relation between infinitely divisible distributions, BFs and CMFs is encoded as follows: for $h: (0, \infty)\mapsto (0, \infty)$ the following statements are equivalent: {\bf (i)} $h$ is CMF and it is infinitely divisible with $h(0+) \leq 1$; and {\bf (ii)} $h = \exp(-{\widehat{\Psi}})$ with ${\widehat{\Psi}}$ being BF \cite[p. 52, Lemma 5.8]{RSchilling10}. Coming back to the relaxation phenomena we remind that the relaxation function $n(t)$ (which provides us with the information on the number of relaxation centers which did not decay during the time $(0,t)$) if calculated from the spectroscopic data appears to be CMF for a vast majority of commonly used phenomenological models \cite{Hanyga1,EDeOliveira11,RGarrappa11,Tomovski,RGarrappa16,KGorska20a,KGorska21}. This fact merged with the just mentioned theorem strongly suggests that characteristic exponents are inextricably linked with investigation of the relaxation processes. Research which sheds light on this problem is the leitmotif of our paper.

We present and discuss a number of arguments which clarify the role played by characteristic exponents in description of the relaxation phenomena, in particular provide the reader their interpretation as memory functions. The methods which we advocate are general and, as recently demonstrated in  \cite{KGorska21a,SW21}, applicable to various phenomenological models of relaxation. In what follows we focus our attention on the excess wings model \cite{RHilfer02b,RHilfer02c} which goes beyond the Jonscher universal relaxation law (URL) \cite{AJonscher92} and is less popular among experimentalists if compared with models of the Havriliak-Negami family. General considerations of Sec. \ref{sec1} show how the characteristic exponent  enters the spectral, relaxation, and memory functions. Also we explain the physical interpretation of ${\widehat{\Psi}}(s)$ and demonstrate that required equivalence of Eqs. \eqref{19/02-2} and \eqref{19/02-1} inevitably leads to the concept of the Sonine pair which non--negligible role in theoretical studies of relaxation, viscoelasticity and anomalous diffusion was recently noticed, analysed and developed \cite{AHanyga20, AGiusti20, AGiusti20a}. Starting from Sec. \ref{sec2} we investigate the excess wings model. Using its spectral function we recover suitable characteristic exponent whose knowledge enables us to find the appropriate relaxation function. In Sec. \ref{sec3} we use the characteristic exponent to introduce two coupled memory kernel functions which form the Sonine pair. Thus we arrive at a pair of evolution equations which involve  either the smearing of the relaxation function or its time derivative and should give the same excess wings relaxation function.  Both equations are solved in Sec. \ref{sec4} where also  requirements demanded from their solutions are checked.  The paper is concluded in Sec. \ref{sec6}.

\section{Characteristic exponents as constitutive elements of the relaxation theory }\label{sec1}

As sygnalized in the Introduction the first step in the construction of stochastic approach to relaxation phenomena, see \cite{AStanislavsky15} and \cite{AStanislavsky17,AStanislavsky19_1} for recent exhaustive reviews, is to assume that they are underpinned by randomization of the characteristic time $\tau$ and that the stochastic processes $U(\tau)$ emerging from such a randomization have non--negative infinitely divisible distributions. In the majority of physically meaningful applications these distributions are realized as heavy tailed $\alpha$-stable L\'evy ones related to various variants of random walks.  The second step, essential for making the method effective in modeling physical applications, is to use the subordination formalism \cite{Bochner} within which the parent process, usually the Debye law dependent on operational time $\tau$, is subordinated by a directing process $\tau(t)$ which links $\tau$ and $t$  in a random relation encoded in a probability density (pdf) $f(\tau,t)$. Intuitively, employing the subordination scheme means to replace a process described in terms of the laboratory clock measured time $t$ by a composed random process governed by an irregular non-decreasing flow of randomized time $\tau$ given by a stochastic process $t\to\tau=\tau(t)$. Physically it is expected that properties of $\tau(t)$ may shed light on the internal structure of the system or provide us with some hints how its macroscopic behaviour is influenced by many-body effects.

According to Eq. \eqref{27/07-1} the probability theory introduces the characteristic exponent $\widehat{\Psi}(s)$ in terms of the mean value of the exponentiated non-negative stochastic process $\exp{(-sU(\tau))}$. Suppose that  $f(\tau,t)$ (which says how to find the system in the operational time $\tau$ if it is in the laboratory time $t$) is also the infinitely divisible pdf of $U(\tau)$.  Then,
\begin{align}\label{17/02-2}
\begin{split}
     \mathcal{L}[f(\tau, t); s] & = \int_{0}^{\infty} \E^{-st} f(\tau, t) \D t \\&=\left\langle \E^{-s U(\tau)}\right\rangle = \E^{-\tau {\widehat{\Psi}}(s)}, \quad \tau > 0.
	 \end{split}
	 \end{align}
Assumption that the process $U(\tau)$ results from $t\to\tau: U(\tau)\le t$ with the pdf $f(\tau,t)$ opens the possibility to ask for the ``inverse'' process $S(t)= {\rm inf}\{ \tau: U(\tau)>t \}$ and its pdf $g(t,\tau)$. The latter may be calculated from the cumulant distribution functions of $U(\tau)$ and $S(t)$ (see e.g. \cite{AStanislavsky15,ChechkinSokolov21}) 
 \begin{equation}\label{28/08-1}
   g(t,\tau)=-\frac{\partial}{\partial\tau}\int_{0}^{\,t} f(\tau, \xi)\D\xi.
	 \end{equation}  
Taking the Laplace transform of Eq. \eqref{28/08-1} and using Eq. \eqref{17/02-2} leads to \cite{L4}
 \begin{equation}\label{28/08-3}
   \widehat{g}(s, \tau) = \frac{\widehat{\Psi}(s)}{s} \E^{-\tau\widehat{\Psi}(s)}.
	 \end{equation}
Alternatively, any non-Debye relaxation process may be seen as summing up effects of multichannel exponential decays with each channel characterized by  some randomly distributed relaxation time $\theta$. Under this assumption the relaxation function $n(t)$ counting the fraction of objects which have survived the decay in the laboratory time interval $(0,t)$ boils down to the weighted average of exponential decays 
 \begin{equation}\label{28/08-4}
  n(t)=\int_{0}^{\infty}e^{-t\theta}\mu(\D\theta),
	 \end{equation}
 where $\mu(\D\theta)$ denotes the probability with which the random relaxation time $\theta$ occurs. In the framework of the subordination approach the same quantity $n(t)$ comes from weighted average of the Debye law expressed in the operational time $\tau$  and the pdf $g(t,\tau)$. Thus  Eq. \eqref{28/08-4} may be rewritten as the  integral decomposition \cite{Fogedby94}      
\begin{equation}\label{28/08-5}
  n(t)=\int_{0}^{\infty}e^{-B({ p})\, \tau}g(t,\tau)\D\tau. 
	 \end{equation}
Using Eq. \eqref{28/08-5} enables us to calculate the response (called also spectral) function defined in the frequency domain as $\widehat{\phi}(\I\!\omega) = {\cal L}^{-1}[-\dot{n}(t); \I\!\omega]$. Because of Eqns. \eqref{28/08-3} and \eqref{28/08-5} it is uniquely expressed in terms  of the characteristic exponent 
 \begin{equation}\label{28/08-6}
  \widehat{\phi}(\I\!\omega) = \frac{1}{1+\widehat{\Psi}(\I\!\omega)/B({ p})} .
	 \end{equation}
Here we point out that Eq. \eqref{28/08-6} explicitly determines the relation between purely phenomenological object  which is the spectral function $\widehat{\phi}$ obtained as a fit to experimental data and $\widehat{\Psi}$, a mathematical quantity one to one  related to the stochastic process being assumed to underlie physical phenomenon under consideration but of origin rather loosely supported by specific physical properties of the system.  To look for physical justification of so far presented construction notice that the relation
\begin{equation}\label{28/08-5a}
  \widehat{n}(s)=\frac{1-\widehat{\phi}(s)}{s}. 
	 \end{equation}
and Eq. \eqref{28/08-6} implies
\begin{equation}\label{28/08-5b}
  \widehat{n}(s)=\frac{s^{-1}}{1+B(p)/\widehat{\Psi}(s)}. 
	 \end{equation}
As recalled in the Introduction the time evolution equations involving memory effects may be obtained by modeling memory effects through the time smearing, either of $\dot{n}(t)$ like it has taken place in Eq. \eqref{19/02-2} or of $r(t, \tau)n(t)$ like has been done in Eq. \eqref{19/02-1}. Doing that we arrive at linear integro-differential equations which without difficulties may be solved in the Laplace domain. The relaxation function which solves Eq. \eqref{19/02-2} in the Laplace domain reads
\begin{equation}\label{1/05-2}
\widehat{n}_{k}(s) = \frac{s^{-1}}{1 + B( { p})/[s~\widehat{k}(s)]}. 
\end{equation}
while for Eq. \eqref{19/02-1} we get
\begin{equation}\label{1/05-1}
\widehat{n}_{M}(s) = \frac{s^{-1}}{1 + B( { p})~ \widehat{M}(s)}, 
\end{equation}
where $\widehat{k}(s) = \mathcal{L}[k(t); s]$ and $\widehat{M}(s) = \mathcal{L}[M(t); s]$. Physical equivalence of the above approaches requires that Eqs. \eqref{1/05-2} and \eqref{1/05-1} describe the same situation, i.e., the memory effects influencing the behaviour  of $\dot{n}(t)$ and $r(t, \tau) n(t)$ should yield the same results for observed properties of $n(t)$. The equality of Eqs. \eqref{1/05-2} and \eqref{1/05-1},  i.e., $\widehat{n}_{k}(s)=\,\widehat{n}_{M}\,= \widehat{n}(s)$, if compared with Eq. \eqref{28/08-5b}, gives 
\begin{equation}\label{1/05-3a}
\widehat{M}(s) = [s~\widehat{k}(s)]^{-1} = [\widehat{\Psi}(s)]^{-1},
 \end{equation}
which merges the deterministic, i.e. evolution equations stemmed, description of the relaxation with its stochastic roots. Consequently, the stochastic nature of relaxations puts rigid restrictions on properties of admissible memory functions, in fact deeply reaching for their analyticity structure. This is because the memory functions form not only the Sonine pair written down in the Laplace domain as
$\widehat{M}(s)\widehat{k}(s) = s^{-1}$ but being directly related to the characteristic exponents and Bernstein functions (both living on the positive semiaxis) may be consistently extended to the complex domain where they fall into special classes of analytic functions, namely the Stieltjes and Nevanlinna-Pick functions \cite{GGripenberg,Berg}.   

\section{The excess wings model}\label{sec2}

The spectral function which corresponds to the simplest version of the excess wings model is
   \begin{equation}\label{29/02-1}
      \widehat{\phi}_{\alpha}(\I\!\omega) = \frac{1 + (\I\!\omega\tau_2)^\alpha}{1 + \I\!\omega\tau_1+ 
			                             (\I\!\omega\tau_2)^\alpha}, \quad \alpha\in(0, 1).
   \end{equation}
It depends on two characteristic times $\tau_1 > 0$ and $\tau_2 > 0$ and so does not fit to the Jonscher's URL \cite{RGarrappa16,AJonscher92,SW16} involving only a single characteristic time $\tau$. Despite this reservation the excess wings model appears useful in analysis of experimental data as it successfully describes the relaxation phenomena in the high frequency regime when the frequency of applied electric field is of the order $10^{5}-10^{10}$ Hz \cite{RHilfer02b,RHilfer02c,PDixon90,PDixon90a,RBrand00}. For $\alpha = 1$ the spectral function $\widehat{\phi}_{\alpha}(\I\!\omega)$  is proportional to the Debye spectral function $\widehat{\phi}_{D}(\I\!\omega) = [1 + (\tau_{1} + \tau_{2}) \I\!\omega]^{-1}$ with the characteristic time $\tau_{1} + \tau_{2}$, i.e., $\widehat{\phi}_{1}(\I\!\omega) = \widehat{\phi}_{D}(\I\!\omega) + \I\!\omega\tau_{2}\widehat{\phi}_{D}(\I\!\omega)$. 

Comparing the spectral function \eqref{29/02-1} with \eqref{28/08-6} we find that the characteristic exponent $\widehat{\Psi}$ formally reads 
   \begin{equation}\label{29/02-3}
      \widehat{\Psi}(\I\!\omega) = \frac{\I\!\omega}{\tau_{2}^{-\alpha} + (\I\!\omega)^{\alpha}}, \quad \alpha\in(0, 1)
   \end{equation}
if we set $B(\tau, {p}) = B(\tau_{1}, \tau_{2}, \alpha) = \tau_{2}^{\alpha}/\tau_{1} = \widetilde{\tau}$.  But some doubt arises: is the construction described in Sec. \ref{sec1} legitimate if we have two characteristic times - which of them, and how, is randomized? To find out properties of $\widehat{\Psi}$ without referring to the L\'evy-Kchintchine formula  consider the function $\widehat{\Psi}(\I\!\omega)$ given by Eq. \eqref{29/02-3} as the function $\widehat{\Psi}(z)$ of a complex variable $z\in\mathbb{C}$. As shown in \cite[Eq. (2.22) {\it et seq.}]{EDeOliveira11} this function satisfies all conditions of \cite[Theorem 2.6]{GGripenberg} or \cite[Theorem]{EDeOliveira11}. It  leads to the crucially important result - namely enables us to represent in an unique way $\widehat{\Psi}(z)$ as the Laplace transform of a non-negative function. Furthermore, restricting the argument $z$ of $\widehat{\Psi}(z)$ to the positive semiaxis, i.e., $z=s>0$, we can identify $\widehat{\Psi}(s)\vert_{s\in\mathbb{R}_{+}}$ as a BF and make use of a plethora of results concerning CMFs and BFs. In the first step notice that for $s>0$ the function $\widehat{\Psi}(s)$ is non-negative while its first derivative 
   \begin{equation*}\label{29/02-4}
      \frac{\D \widehat{\Psi}(s)}{\D s} = \frac{1-\alpha}{\tau_2^{-\alpha} + s^\alpha} 
			        + \frac{\alpha \tau_2^{-\alpha}}{(\tau_2^{-\alpha} + s^\alpha)^2}
   \end{equation*}
is CMF. 
\begin{table}
\begin{center}
\begin{tabular}{c | c | c }
$f(s)$ $\quad$ & $\quad$ CMF $\quad$ & $\quad$ BF $\quad$\\ \hline 
$(s + b)^{\,\mu}$, $b \geq 0$ & $\mu \leq 0$ & $\mu\in(0, 1)$  \\ 
$(s^{\nu} + b)^{\,\mu}$, $b > 0$ & $\mu \leq 0$ and $\nu\in(0, 1]$ & $\mu\in(0, 1)$ and $\nu\in(0, 1)$
\end{tabular}
\caption{\label{tab1} Examples of CMF and BF.}
\end{center}
\end{table}
 Indeed, from Tab. \ref{tab1} we see that for non-negative $\tau_{2}$ and $\alpha\in(0, 1)$ this expression is a convex sum of CMFs and  hence it is CMF as well. Thus, the characteristic exponent  $\widehat{\Psi}(s)$ itself is BF. This is the result which we do need and which for the case under consideration is by no means obvious from the stochastic point of view since we lack the information concerning the infinite divisibility of underlying stochastic process. Needed result, which obviously confirms infinite divisibility, is obtained from the completely different sources, namely from the phenomenology merged with mathematical analysis. We would also like to remark that within the stochastic approach we deal with functions of real variables  exemplified by those being CMFs and BFs. Starting from the spectral function treated as a complex function of the complex variable we avoid the path marked out by principles of the stochastic approach. Equipped with tools of the complex analysis we can leave aside the probability rooted description of relaxation phenomena and may understand much better results not once or twice hidden behind paradigms of the real functions approach \cite{KGorska21}.   

The relaxation function $n(t)$ Eq. \eqref{28/08-5} with substituted Eq. \eqref{28/08-3} reads
   \begin{align}\label{29/02-5}
         n(t) & = \int_{0}^{\infty}\!\! \E^{-\xi \widetilde{\tau}} \mathcal{L}^{-1}\!
				          \left[\frac{1}{\tau_{2}^{-\alpha} 
							  + s^{\alpha}} \exp\Big(- \frac{\xi s}{\tau_{2}^{-\alpha} + s^{\alpha}}\Big) ; t\right] \D\xi 
							  \nonumber \\
							  &= \mathcal{L}^{-1}[(s + s^{\alpha}\widetilde{\tau} + \tau_{1}^{-1})^{-1}; t], 
   \end{align}
where $\widetilde{\tau}=\tau_{2}^{\alpha}/\tau_{1}$ was introduced a few lines above. To get Eq. \eqref{29/02-5} we changed the order of integration over $\xi\in[0, \infty)$ in the inverse Laplace transform which reduced the integral over $\xi$ to the elementary one. The inverse Laplace transform in the lower line of Eq. \eqref{29/02-5} can be calculated by virtue of the formula 
\cite[p. 10, Eq. (1.38)]{TSandev19} 
   \begin{multline*}
	    \mathcal L^{-1}\Big[\dfrac{s^{-\beta}}{1+\lambda_1s^{-\alpha_1} + \lambda_2s^{-\alpha_2}}; t\Big] \\
			               = t^{\,\beta-1}\, E_{(\alpha_1, \alpha_2), \,\beta}\big(-\lambda_1 t^{\alpha_1}, 
					      - \lambda_2 t^{\alpha_2} \big), 
	 \end{multline*}
from which we get  
   \begin{equation} \label{1/03-1}
      n(t) = E_{(1, 1-\alpha), 1}(- t/\tau_{1}, -\widetilde{\tau} t^{1-\alpha}).
   \end{equation}
As shown in \cite{RRNigmatulin16} this function, known as the binomial (multivariable) Mittag-Leffler function,  is defined by the double power series
 \begin{equation}\label{1/03-2}
      E_{(\alpha_{1}, \alpha_{2}), \,\beta}(x, y) =  \sum_{k \geq 0} 
			  \sum_{\underset{l_1+l_2=k}{l_{1}, l_2 \geq 0}} \frac{k!}{l_{1}!\, l_{2}!}\, 
				\frac{x^{l_{1}} y^{l_{2}}}{\Gamma(\beta + \alpha_{1} l_{1} + \alpha_{2} l_{2})}, 
   \end{equation}
,$x, y \in \mathbb{R}$, and is non-negative for $\lambda_{1}, \lambda_{2} \geq 0$, $\beta \in (0, 1)$, and $\alpha_{1}, \alpha_{2} \geq \beta - 1$. Thus, $n(t)$ given by Eq. \eqref{1/03-1} is non-negative for $\alpha\in(0, 1)$. Notice that in Eq. \eqref{1/03-2} the infinite sum over $k$ is followed by sums over $l_{1}$ and $l_{2}$ constrained by $l_{1} + l_{2} = k$. As a consequence the double sum in $l_{1}$ and $l_{2}$ can be represented two-fold: {\bf (a)}  $l_{1} = 0, 1, \ldots, k$ and $l_{2} = k - l_{1}$ or {\bf (b)} $l_{2} = 0, 1, \ldots k$ and 
$l_{1} = k - l_{2}$. Without loss of generality we consider the case {\bf (a)}. In such a case Eq. \eqref{1/03-1} becomes
   \begin{align} \label{2/03-0}
      n(t) & = \sum_{k\geq 0} \sum_{l_{1} = 0}^{k} \binom{k}{l_{1}} 
			       \frac{(-t/\tau_{1})^{k} (\tau_{2}/t)^{\alpha(k-l_{1})}}{\Gamma[1+k - \alpha(k-l_{1})]}\\ \nonumber
	    & = \sum_{k\geq 0} (- \widetilde{\tau} t^{1-\alpha})^{k} \sum_{l_{1}=0}^{k} \binom{k}{l_{1}}	 \frac{(t/\tau_{2})^{\alpha l_{1}}}{\Gamma[1 + (1-\alpha)k + \alpha l_{1}]}.      
   \end{align}
   Using the definition of Mittag-Leffler polynomials \eqref{24/03-a3} we get
   \begin{equation}\label{10/08-1}
   n(t) = \sum_{k\geq 0} (- \widetilde{\tau} t^{1-\alpha})^{k} E_{\alpha, (1-\alpha)k + 1}^{-k}[-(t/\tau_{2})^{\alpha}].
   \end{equation}
The same expression as in Eq. \eqref{2/03-0} will be obtained if we change $\sum_{k \geq 0}\sum_{l_1=0}^k$ into 
$\sum_{l_1\geq 0} \sum_{k\geq l_1}$. Making this change and setting $r = k - l_{1}$ we transform the series and the sum sitting inside Eq. \eqref{2/03-0} into two independent series
\begin{equation}\label{6/08-1}
n(t) = \sum_{l_{1} \geq 0} \sum_{r \geq 0} \frac{(l_{1} + r)!}{r! l_{1}!} \frac{(-\widetilde{\tau} t^{1-\alpha})^{r} (-t/\tau_{1})^{l_{1}}}{\Gamma[1 + (1-\alpha) r + l_{1}]}.
\end{equation}
Treating once the series over $l_{1}$ and another time the series over $r$ as the definition \eqref{24/03-a2} of the three parameter Mittag--Leffler (or Prabhakar) function \cite{L5} we express Eq. \eqref{6/08-1} in two equivalent forms, namely
   \begin{align}\label{2/03-1a}
      n(t) & = \sum_{l_{1}\geq 0} (-t/\tau_{1})^{l_{1}} E^{1+l_{1}}_{1-\alpha, 1 + l_{1}}(-\widetilde{\tau} t^{1-\alpha}) \\
							 \label{2/03-1b}
           & = \sum_{r\geq 0} (-\widetilde{\tau} t^{1-\alpha})^r
					     E^{1+r}_{1, (1-\alpha)r + 1}(-t/\tau_{1})\,.
   \end{align}
    Calculations made for {\bf (a)} can be repeated for {\bf (b)} with $l_{2}$ written instead of $l_{1}$; thus, $r = k - l_{2}$. The formulae \eqref{2/03-1a} and \eqref{2/03-1b} reproduce the relations \cite[Eqs. (3.73), (3.71)]{RGarrappa16} up to the multiplicative constant $1/\tau_1$. Moreover, we conclude that $n(0+) = 1$.
  
\section{Evolution equation}\label{sec3}

\subsection{Smearing of $r(t, \tau) n(t)$}
First we check what equation is satisfied by $n(t)$. For that purpose we take Eq. \eqref{10/08-1}. In 
\cite[Theorem 2.3.1. on p. 93]{AMMathai08}, i.e. $z E_{\mu, \nu}^\gamma(z) = E_{\mu, \nu-\mu}^\gamma(z) - E_{\mu, \nu-\mu}^{\gamma-1}(z)$, we set $\mu = \alpha$, $\nu = (1-\alpha)k + 1$, and $\gamma = -k$. That allows us to rewrite Eq. \eqref{10/08-1} in the form
\begin{align}
&n(t) = -\tau_{1} \sum_{k=0}^{\infty} (-\widetilde{\tau}\,)^{\,k+1} t^{(1-\alpha)(k+1) -1} \nonumber\\ & \times \left\{E^{-(1+k)}_{\alpha, (1-\alpha)(1+k)}[-(t/\tau_{2})^{\alpha}] - E^{-k}_{\alpha, (1-\alpha)(1+k)}[-(t/\tau_{2})^{\alpha}] \right\} \nonumber \\
& = -\tau_{1} \left\{ \sum_{r=1}^{\infty} (-\widetilde{\tau}\,)^{\,r} t^{(1-\alpha) r - 1} E^{\,-r}_{\alpha, (1-\alpha) r}[-(t/\tau_{2})^{\alpha}] \right. \nonumber \\ \label{10/08-2}
& \left. + \widetilde{\tau} \sum_{k=0}^{\infty} (-\widetilde{\tau}\,)^{k} t^{(1-\alpha) k - \alpha} E^{-k}_{\alpha, (1-\alpha)k + 1 - \alpha}[-(t/\tau_{2})^{\alpha}]\right\} 
\end{align}
where we change the summation index in the first series by setting $k+1 = r$. From Eqs. \eqref{10/08-3} and \eqref{10/08-4} it comes out that Eq. \eqref{10/08-2} can be expressed as
\begin{equation}\label{10/08-5}
- \frac{1}{\tau_{1}}\, n(t) = \Big[\frac{\D}{\D t} + \widetilde{\tau} D_{t}^{\alpha}\Big]n(t), \quad \alpha\in(0, 1),
\end{equation}
where $({D_{x}^{\alpha}} f)(x) = (\frac{\D}{\D x}I_{0}^{1-\alpha}f)(x)$ is the fractional derivative in the Riemann-Liouville sense for $\alpha \in (0, 1)$ whereas $(I_0^{\,\nu} f)(x)$ (given by Eq. \eqref{10/08-10}) is the Riemann-Liouville fractional integral for $\nu\in(0, 1)$. We point out that the time operator in square bracket of \eqref{10/08-5} is equivalent to \cite[Eq. (5) for $\beta = 0$]{RHilfer17}.  
Acting with $I_0^{\,1}$ on both sides of Eq. \eqref{10/08-5} we get
\begin{equation*}
-\frac{1}{\tau_{1}} (I_{0}^{1}n)(t) = n(t) - 1 + \widetilde{\tau}\,(I^{1-\alpha}_{0} n)(t)
\end{equation*}
represented also in the form 
  \begin{equation*}
  n(t) = 1 - \dfrac1{\tau_1}\,(I^{1}_{0}\, n)(t) - \widetilde{\tau}\,(I^{1-\alpha}_{0} n)(t).
  \end{equation*}
That leads to Eq. \eqref{19/02-1} with $B({\tau}_{1},{\tau}_{2}, {\alpha}) = \widetilde{\tau} = \tau_{2}^{\alpha}/\tau_{1}$ and
 \begin{equation}\label{10/08-11}
 M(t) = \tau_{2}^{-\alpha} + \frac{t^{-\alpha}}{\Gamma(1 - \alpha)}, 
 \end{equation}
which is interpreted as power-like smearing of $r(t, \tau) n(t)$ in Eq. \eqref{6/08-2}. The related Laplace transform becomes
\begin{equation}\label{10/08-12}
\widehat{M}(s) = \frac{\tau_{2}^{-\alpha} + s^{\alpha}}{s}.
\end{equation}
From the above and Eq. \eqref{1/05-3a} we restore the characteristic exponent described by Eq. \eqref{29/02-3}.

\subsection{Coupled memories}

The explicit form of the characteristic exponent  $\widehat{\Psi}(s)$ enables us to find  
memories $M(t)$ and $k(t)$ responsible for the time smearing of Eq. \eqref{6/08-2}. Recall that the memory $M(t)$ reflects the smearing of $n(t)$ whereas $k(t)$ is related to the smearing of the time derivative $\dot n(t)$ and that the memory $M(t)$ and its Laplace form are given by Eqs. \eqref{10/08-11} and \eqref{10/08-12}.  Using the coupled pair $\widehat{M}(s) \widehat{k}(s) = 1/s$ we find that $\widehat{k}(s)$ and its Laplace form $k(t)$ yield
   \begin{equation}\label{4/03-1}
    			\widehat{k}(s) = (\tau_{2}^{-\alpha} + s^{\alpha})^{-1} \quad \text{and} \quad k(t) = t^{\alpha-1} E_{\alpha, \alpha}[-(t/\tau_{2})^{\alpha}].
   \end{equation}
The singularity of $M(t)$ and $k(t)$ at $t = 0$ is controlled by the parameter $\alpha$. In the example quoted just below Corollary 4.1 in  \cite{AHanyga20} it is pointed out that $M(t)$ and $k(t)$ are the so--called Sonine functions and the coupled pair $(k, M)$ is the Sonine pair \cite[pp. 213--4]{AHanyga20}; at a moment we conclude that they are only Sonine functions $k$ and $M$. Such functions are locally integrable non--decreasing functions which satisfy  
   \[ \sigma(t) \to \infty, \quad t\sigma(t) \to 0, \quad \text{for} \quad t \to 0;\quad\sigma \in \{M, k\}. \]
Thus, \cite[Theorem 3.1]{AHanyga20} is revealed. According to the philosophy of the coupled me-\\mories $M(t)$ is linked to Eq. \eqref{10/08-5} and $k(t)$ to
  \begin{equation}\label{6/03-5}
      \int_{0}^{t} (t - \xi)^{\alpha -1} E_{\alpha, \alpha}\Big[-\Big(\frac{t-\xi}
			             {\tau_{2}}\Big)^{\alpha}\Big]~ \dot n(\xi) \D\xi = - \widetilde{\tau}\, n(t).
   \end{equation}
Hence, the smearing of the relaxation function $n(t)$ can be changed into the smearing of its first 
time derivative $\dot{n}(t)$ like it is done in Eq. \eqref{6/03-5}. 

\section{The series form of solutions to \eqref{6/03-5}}\label{sec4}

General conditions of solvability  Eq.\eqref{6/03-5} are precised in \cite[Theorem 2]{ANKochubei11}. 
It guarantees the uniqueness of the solution, its continuity, differentiability, and completely 
monotone character on $(0, \infty)$. From Eq. \eqref{4/03-1} the asymptotics of $\widehat{k}(s)$
turns out to be 
   \begin{align*}
      \begin{split}
         \widehat{k}(s) \to \tau_{2}^{\alpha}, \quad s~ \widehat{k}(s) \to 0, 
				    \quad \text{for} \quad s\to 0, \\
         \widehat{k}(s) \to 0, \quad s~ \widehat{k}(s) \to \infty, 
				    \quad \text{for} \quad s\to \infty,
      \end{split}
   \end{align*}
so we reconstruct the conditions listed in \cite[Theorem 2]{ANKochubei11} except of the first of them: 
$\widehat{k}(s)$ does not tend to infinity with $s\to 0$ but to the constant $\tau_{2}^{\alpha}$ instead. 
This clearly suggests the existence of a solution to \eqref{6/03-5} which differs from \eqref{1/03-1}.

Looking for the solution of \eqref{6/03-5} we apply \cite[Eq. (5)]{KGorska20}:
   \begin{equation*} 
   n(t) = \mathcal{L}^{-1} [\widehat{k}(s)/(s~ \widehat{k}(s)+\widetilde{\tau}); t ],
   \end{equation*}
in which we extract from denominator either $s~ \widehat{k}(s)$ or $\widetilde{\tau}$. This extraction procedure enables us to infer two kinds of formulae. To derive them we employ the series form of $(1+x)^{-1} = \sum_{r\geq 0} (-x)^{r}$ for $|x| < 1$, where we take either $x = \widetilde{\tau} [s~ \widehat{k}(s)]^{-1}$ or $x = s~\widehat{k}(s)/\widetilde{\tau}$, getting the series form of solutions $n_{\alpha}(\tau_1, \tau_2; t)$ and $\widetilde n_{\alpha}(\tau_1, \tau_2; t)$, respectively. Clearly   
   \begin{align*}
      n_{\alpha}(\tau_1, \tau_2; t) & = \sum_{r\geq 0} (- \widetilde{\tau})^r 
			   \mathcal{L}^{-1}\Big[\dfrac{(\tau_{2}^{-\alpha} + s^{\alpha})^r}{s^{1+r}}; t\Big], \\
      \widetilde{n}_\alpha(\tau_1, \tau_2; t) & = \frac1{\widetilde{\tau}} \sum_{r\geq 0} 
			   (-\widetilde{\tau})^{-r} \mathcal{L}^{-1}\Big[\dfrac{s^r}{(\tau_{2}^{-\alpha} 
			 + s^{\alpha})^{1+r}}; t\Big].
   \end{align*}
We point out that these formulae are equivalent to those obtained in \cite[Eqs. (1.4), (3.1)]
{KGorska19} or \cite[Eqs. (6), (7)]{KGorska20}). The inverse Laplace transforms present in 
$n_{\alpha}(\tau_{1}, \tau_{2}; t)$ and $\widetilde{n}_{\alpha}(\tau_{1}, \tau_{2}; t)$ are 
calculated applying the technique of \cite[Eq. (2.5)]{TRPrabhakar69} exhibited in Eq. \eqref{24/03-a1}. 
For $\alpha\in(0, 1)$ we have
   \begin{equation} \label{16/03-4}
      n_\alpha(\tau_{1}, \tau_{2}; t) = \sum_{r\geq 0} (-\widetilde{\tau})^{r}\, t^{(1-\alpha)r} E_{\alpha, (1-\alpha)r + 1}^{-r}[-(t/\tau_2)^\alpha], 
   \end{equation}
while for $\alpha>1$
   \begin{equation} \label{16/03-5}
      \widetilde{n}_{\alpha}(\tau_{1}, \tau_{2}; t) = \frac{1}{\widetilde{\tau}} \sum_{r\geq 0} 
			               (-\widetilde{\tau})^{-r}\, t^{(\alpha -1)r}  E_{\alpha, (\alpha - 1)r + 1}^{\,r}[-(t/\tau_2)^\alpha].
   \end{equation}
   (Notice that Eq. \eqref{16/03-4} is the same as Eq. \eqref{10/08-1}.) Both calculation procedures are legitimate because the Mittag--Leffler functions setting in the series, either \eqref{16/03-4} or \eqref{16/03-5}, are well defined for all $r\in \mathbb N_0$ as depending on the parameters $(1-\alpha)r + 1>0; \alpha \in (0, 1)$, and $(\alpha-1)r + 1>0; \alpha>1$, respectively. We point out that the Laplace transforms in both solutions yield to $(s + s^\alpha\, \widetilde{\tau} + \tau_1^{-1})^{-1}$, once for $\alpha\in(0, 1)$ and in turn for $\alpha > 1$. In the excess 
wings model we have $\alpha\in (0, 1)$ so $n_\alpha(\tau_1, \tau_2; t)$ is the correct solution for 
that range of $\alpha$. In the case of non-negative integer $n \in \mathbb N_0$, the 
expression $E_{\alpha, \nu}^{-n}(z)$ becomes the Mittag--Leffler polynomial of degree 
$\deg\big(E_{\alpha, \nu}^{-n}\big) = n$ which basic properties are quoted in \ref{appA}. 

For $\alpha = 1$ the solutions coincide, $\widetilde{n}_1(\tau_1, \tau_2; t) = n_1(\tau_1, \tau_2; t)$, taking the exponential decay form
   \begin{equation*}
   n_1(\tau_1, \tau_2; t) = \frac{\tau_1}{\tau_1 + \tau_2} \exp\Big(-\frac{t}{\tau_1 + \tau_2}\Big),
   \end{equation*}
viz., \ref{appC}. The equality of $n_\alpha(\tau_1, \tau_2; t)$ and $\widetilde{n}_\alpha(\tau_1, \tau_2; t)$ for $\alpha\neq 1$ can be established in the limit case of large $\tau_2$ for which the uniqueness conditions of the initial Cauchy problem \eqref{6/03-5} and $n(0+)=1$, given in \cite{ANKochubei11}, are satisfied. The limit of large $\tau_2$ means, assuming $t$ be fixed, 
that the three parameter Mittag-Leffler function and the Mittag-Leffler polynomial are considered for small values of their arguments being in both cases equal to $t/\tau_2$. The asymptotic behaviour either of the Mittag--Leffler function or of the associated $\deg(E^{-n}_{\alpha, \,\beta})$ 
Mittag-Leffler polynomial coincide for a small values of argument: 
   \begin{equation*}
      E^{\gamma}_{\alpha, \,\beta}(x)\quad \text{or} \quad E^{-n}_{\alpha, \,\beta}(x) \sim [\Gamma(\beta)]^{-1}, 
	       \qquad x \to 0\,.
	  \end{equation*} 
Accordingly, when $\tau_2$ is growing and $t$ remains fixed, we deduce 
   \[ n_\alpha(\tau_1, \tau_2; t) \propto \sum_{r\geq 0} \frac{(-\widetilde{\tau} t^{1-\alpha})^r}
	      {\Gamma[1 + (1-\alpha)r]} = E_{1-\alpha}(-\widetilde{\tau} t^{1-\alpha}). \]
Using the reciprocal arguments property \cite[Eq. (4.8.5)]{RGorenflo14} 
   \begin{equation*}
	    E_{-\nu}(z) + E_{\nu}(z^{-1}) = 1, \qquad \nu>0,\,z \in \mathbb C \setminus \{0\},
	 \end{equation*}
we have 
   \begin{equation*}
      n_\alpha(\tau_1, \tau_2; t) \propto 1 - E_{\alpha-1}[-(\widetilde{\tau} t^{1-\alpha})^{-1}]  = \sum_{r\geq 1} \frac{(-\widetilde{\tau} t^{1-\alpha})^{-r}}{\Gamma\big(1 - (1-\alpha) r\big)}, 
   \end{equation*}
which is the asymptotics of $\widetilde{n}_\alpha(\tau_1, \tau_2; t)$ for $\tau_2 \to \infty$. Thus, we 
infer that $\tau_2 \to \infty$ means $\widehat{k}(s) \to \infty$ which confirms the equality 
$n_\alpha(\tau_1,\tau_2; t) = \widetilde{n}_\alpha(\tau_1, \tau_2; t)$, by bearing in mind the 
uniqueness of solution guaranteed by \cite[Theorem 2]{ANKochubei11}.

After some routine, but long and a little boring calculations employing definitions of the Mittag-Leffler polynomials and the three parameter Mittag-Leffler function, we get that the solutions $n_{\alpha}(\tau_1, \tau_2; t)$ and $\widetilde{n}_{\alpha}(\tau_1, \tau_2; t)$ can be presented in the form of binomial Mittag-Leffler functions:
   \begin{align}\label{17/03-1}
      n_{\alpha}(\tau_1, \tau_2; t) & = E_{(1, 1-\alpha), 1}(- t/\tau_1, 
			        -\widetilde{\tau} t^{1-\alpha}), \\ \label{17/03-2}
      \widetilde{n}_{\alpha}(\tau_1, \tau_2; t) & = t^{\alpha-1}\, 
			 E_{(\alpha, \alpha - 1), \alpha}[-(t/\tau_{2})^{\alpha}, -
							t^{\,\alpha-1}/\,\widetilde{\tau}\,]/\,\widetilde{\tau}.
   \end{align}
We remind that the first of these results holds for $\alpha\in(0, 1)$ while the second one when $\alpha > 1$. In turn, the equality $n_1(\tau_1, \tau_2; t) = \widetilde{n}_1(\tau_1, \tau_2; t)$ yields $E_{(1, 1), 1}(-t/\tau_{1}, - \widetilde\tau) = 1/\widetilde{\tau} \,E_{(1, 1), 1}(-t/\tau_{2}, - 1/\,\widetilde{\tau}\,)$. Reformulation of the Eqs. \eqref{16/03-4} and \eqref{16/03-5} by Eqs. \eqref{17/03-1} and \eqref{17/03-2} involves plenty of technical details, first of all concerning transformations of finite and infinite sums. All this goes beyond the presented exposition and is shifted to \ref{appB}. Finally, we remark that $n_\alpha(\tau_1, \tau_2; t)$ in \eqref{17/03-1} can be expressed as \eqref{2/03-1a} and \eqref{2/03-1b} or the formulae \cite[Eqs. (3.71), (3.73)]{RGarrappa16}, whereas \eqref{17/03-2} coincides with \cite[Eq. (3.72)]{RGarrappa16}.

\section{Conclusions} \label{sec6}
We have shown that the kinetic equations \eqref{19/02-2} and \eqref{19/02-1} assumed to govern the relaxation phenomena and stemmed from the time smearing of either LHS or RHS in non-Debye evolution equation $\dot{n}(t) = -r(t, \tau) n(t)$ determine their stochastic interpretation.  The crucial role in the presented approach is played by the characteristic exponent $\widehat{\Psi}$ which provides us with a bridge connecting kinetic equations and stochastic methods. Moreover, for a large set of relaxing systems  
$\widehat{\Psi}$ obeys well-defined properties which put it in the class of Bernstein functions and open new ways to push forward  mathematical and physical understanding of the relaxation phenomena. To illustrate our methods we went beyond the family of the Havriliak-Negami models and considered the excess wings model of relaxation. We identified the characteristic exponent related to it and derived and solved kinetic equations which reflect two ways of introducing the memory effects - the time smearing of  $\dot{n}(t)$ or $r(t, \tau) n(t)$ reflected in Eqns. \eqref{19/02-2} and \eqref{19/02-1}, respectively. Natural assumption that both approaches lead to the same physical results allowed us to claim that the memory functions, $M(t)$ and $k(t)$, responsible for both variants of smearing, form the Sonine pair, i.e., their transforms to the Laplace domain satisfy $\widehat{M}(s)\widehat{k}(s)=1/s$. Results of the paper complete and, in a sense, unify so-called deterministic and stochastic processes based investigations of the non-Debye relaxation phenomena. We show that both these approaches are not only mutually related but realize a correspondence principle which joins different, but in fact equivalent views on the same physical problem.      

\section*{Acknowledgments}

K.G. and A.H. have been supported by the Polish National Center for Science (NCN) research grant OPUS12 no. UMO-2016/23/B/ST3/01714. K. G. acknowledges also support under the project Preludium Bis 2 no. UMO-2020/39/O/ST2/01563 awarded by the NCN and NAWA (Polish National Agency For Academic Exchange). The research of T.K.P. has been supported in part by the University of Rijeka, Croatia, under the project {\tt uniri-pr-prirod-19-16}. 

 \section*{Conflict of interest}

The authors declare that they have no conflict of interest.

\appendix

\section{Three parameter Mittag-Leffler function and Mittag-Leffler polynomials}\label{appA}

The three parameter Mittag-Leffler function is defined through the power series 
\cite[p. 97, Eq. (5.1.1)]{RGorenflo14}
   \begin{equation}\label{24/03-a2}
      E_{\alpha, \nu}^{\mu}(x) = \sum_{r\geq 0} \frac{(\mu)_{r} x^{r}}{r! \Gamma(\nu + \alpha r)},
   \end{equation} 
where $\RE(\alpha), \RE(\nu), \RE(\mu) > 0$ and $x \in \mathbb{R}$, $(\mu)_{r}$ denotes 
the familar Pochhammer symbol (raising factorial) equal to 
$\Gamma(\mu + r)/\Gamma(\mu) = \mu (\mu + 1)\ldots (\mu + r-1); r \in \mathbb N_0$. The Pochhammer 
symbol for $\mu = 1$ is equal to $r!$ and Eq. \eqref{24/03-a2} depends on  two parameters $\alpha$ and $\nu$ only. This case is named the two parameter Mittag-Leffler (Wiman) function and it is quoted as 
$E_{\alpha, \nu}(x) = E^1_{\alpha, \nu}(x)$. For $\mu = \nu = 1$ Eq. \eqref{24/03-a2} reduces to the one parameter (standard) Mittag-Leffler function $E_{\alpha}(x) = E_{\alpha, 1}^1(x)$. The Laplace transform of 
$t^{\nu - 1}E_{\alpha, \nu}^{\mu}(\lambda t^{\alpha})$ equals
   \begin{equation}\label{24/03-a1}
      \mathcal{L}[t^{\nu - 1} E_{\alpha, \nu}^\mu(\lambda t^\alpha); s] = 
			         s^{-\nu}(1-\lambda s^{-\alpha})^{-\mu}\,
   \end{equation} 
for $\RE(\nu), \RE(s) > 0$, $|s| > |\lambda|^{1/\RE(\alpha)}$ \cite{TRPrabhakar69}. Derivatives of the three parameter Mittag-Leffler function read
\begin{equation}\label{10/08-3}
x^{\,\nu - 1} E_{\mu, \nu}^{\gamma}(a x^{\,\mu}) = \frac{\D}{\D x}[x^{\nu} E_{\mu, \nu - 1}^{\gamma}(a x^{\,\mu})]
\end{equation}
and
\begin{equation}\label{10/08-4}
x^{\,\nu - \alpha} E_{\mu, \nu - \alpha}^{\gamma}(a x^{\mu}) = D_{x}^{\,\alpha} [x^{\nu} E_{\mu, \nu - 1}^{\gamma}(a x^{\,\mu})],
\end{equation}
where $({D_{x}^{\alpha}} f)(x) = (\frac{\D}{\D x}I_{0}^{1-\alpha}f)(x)$ is the fractional derivative in the Riemann-Liouville sense for $\alpha \in (0, 1)$ and  
\begin{equation}\label{10/08-10}
(I_0^{\,\nu} f)(x) = \dfrac1{\Gamma(\nu)}\, \int_{0}^{x}(x-\xi)^{\nu-1}f(\xi) \D\xi\,, 
	                     \quad \nu \in(0, 1],
\end{equation}	                     
stands for the Riemann-Liouville fractional integral. 

The Mittag-Leffler polynomials occur when the upper parameter in Eq. \eqref{24/03-a2} is a negative integer, i.e., $\gamma =-n, n\in \mathbb N_{0}$. From the definition of the Pochhammer symbol all terms in Eq. \eqref{24/03-a2}  
vanish when the upper parameter $\gamma < -n$ and the series terminates leading to
   \begin{equation}\label{24/03-a3}
      E_{\alpha, 1 + c}^{-n}(x) = \sum_{r=0}^n \binom{n}{r} \frac{(-x)^{r}}{\Gamma(1 + c + \alpha r)}, 
			                          \qquad \alpha, c > 0.  
   \end{equation}
These objects are related to the Konhauser polynomials $Z^c_n(x; k)$ 
\cite[p. 304, Eq. (5)]{JDEKonhauser67} defined {\it via} the formula 
   \[ Z^c_n(x; k) = \dfrac{\Gamma(kn+c+1)}{n!} \sum_{j=0}^n \binom{n}{j} \dfrac{(-x^k)^j}
	                  {\Gamma(kn+c+1)}, \]
where $c>-1$. The latter extend the generalized (associated) Laguerre polynomials 
$L_n^{(c)}(x^k) = Z^c_n(x; k)$ (for the {latter} see below in Appendix B), \cite{KGorska20,HMSrivastava82}. 
The connection formula between Mittag--Leffler and Konhauser polynomials reads \cite[p. 633, Eq. (7)]{Ozar} 
   \[ E_{\alpha, c+1}^{-n}(x^\alpha) = \dfrac{\Gamma(\alpha n+c+1)}{n!}~ Z^c_n(x; \alpha)\,.\] 

\section{The proof of $n_1(\tau_1, \tau_1; t) = \widetilde{n}_1(\tau_1, \tau_1; t)$} \label{appC}

The equality of $n_1(\tau_1, \tau_1; t)$ and $\widetilde{n}_1(\tau_1, \tau_1; t)$ we can established using 
the following three facts: 
   \begin{enumerate}
	    \item[\bf 1.] $E_{1, 1}^{-n}(x) = L_{n}(x)$ where $L_n(x)$ signifies the $n$th Laguerre polynomial; 
			\item[\bf 2.] the generating function for generalized (associated) Laguerre polynomials 
			$L_n^{(\alpha)}(x); \quad L_n^{(0)}(x) \equiv L_n(x)$, which Laplace transform we use, reads  
			\cite[Eq. (5.11.2.1)]{APPrudnikov-v2} 
         \[ \sum_{k\geq 0} t^{k} L_{k}^\alpha(x) = \dfrac1{(1 - t)^{1+\alpha}}\,
	    \exp\Big( \frac{t x}{t-1}\Big) \] 
			for all $|t|<1$. For another generating functions see for instance Ref. 
			\cite{Chat};
			\item[\bf 3.] $E_{1, 1}^{\,r}(-x) = \E^{-x} E_{1, 1}^{-(r-1)}(x)$ which is Kummer's 
			first transformation formula for the confluent hypergeometric function ${}_1F_1$, 
			namely ${}_1F_1(r; 1; -x) = E_{1, 1}^{\,r}(x)$. 
	 \end{enumerate}

\section{Derivation of equations \eqref{17/03-1} and \eqref{17/03-2}}\label{appB}

Substituting the Mittag-Leffler polynomial's expression in Eq. \eqref{16/03-4} after some algebra we 
conclude that
   \[ n_1(\tau_1, \tau_2; t) = \sum_{r \geq 0} \sum_{j=0}^r \binom{r}{j} 
	                \frac{\Big(-\dfrac{\widetilde{\tau}}{t^{\alpha-1}}\Big)^{r-j} 
									\Big(-\dfrac{t}{\tau_1}\Big)^j}{\Gamma\big(j + (1-\alpha)(r-j) + 1\big)}. \]
Setting $j = l_1$ and $r-j=l_1$ we can rewrite the right-hand side above as 
   \[ n_1(\tau_1, \tau_2; t) = \sum_{r \geq 0} {\sum_{\underset{l_1+l_2=r}{l_1, l_2 \geq 0}}} 
	                \frac{r!}{l_1! l_2!} \frac{\Big(-\dfrac{\widetilde{\tau}}{t^{\alpha-1}}\Big)^{l_2} 
									\Big(-\dfrac{t}{\tau_1}\Big)^{l_1}}{\Gamma\big(1 + l_1 + (1-\alpha)l_2\big)}. \]
Comparison with Eq. \eqref{1/03-2} gives Eq. \eqref{17/03-1}. Analogous calculation can be 
done for Eq. \eqref{17/03-2}; during these computations we use the series form of the three 
parameter Mittag-Leffler function.


\begin{thebibliography}{99}

\bibitem{RSAnderssen02} 
Anderssen RS, Loy RJ. Completely monotone fading memory relaxation moduli. Bull Austral Math Soc 2002; 65:449 

\bibitem{RSAnderssen02a} 
Anderssen RS, Loy RJ. Rheological implications of completely monotone fading memory. J Rheol 2002; 46:1459 

\bibitem{Berg} Berg C. Stieltjes-Pick-Bernstein-Schoenberg and their connection to completely monotonicity. In: Mateu J and Porcu E, editors. Positive Define Functions: From Schoenberg to Space-Time challenges. Dep. Math. of Univ. Jaume I, Castellon; 2008

\bibitem{Bochner} Bochner S. Harmonic Analysis and the Theory of Probability. Univ. of California Press, Berkeley/Los Angeles; 1955

\bibitem{RBrand00} Brand R, Lunkenheimer P, Schneider U, Loidl A. Excess wing in the dielectric loss of glass-forming ethanol: A relaxation process. Phys Rev B 2000; 62:8878 

\bibitem{EDeOliveira11} 
Capelas de Oliveira E, Mainardi F, Vaz Jr J. Models based on Mittag-Leffler functions for anomalous relaxation in dielectrics. Eur Phys J Special Topics 2011; 193:161 

\bibitem{Chat} Chatterjea SK. On a generating function of Laguerre polynomials. Boll Un Mat Ital 1962; 17:179 

\bibitem{ChechkinSokolov21} Chechkin AV, Sokolov IM. On relation between generalized diffusion and subordination schemes. Phys Rev E 2021; 103:032133 

\bibitem{PDixon90} Dixon PK. Specific-heat spectroscopy and dielectric susceptibility measurements of $salol$ at the glass transition. Phys Rev B 1990; 42:8179 

\bibitem{PDixon90a} Dixon PK, Wu L, Nagel SR, Williams BD, Carini JP. Scaling in the relaxation of supercooled liquids. Phys Rev Lett 1990; 65:1108 

\bibitem{Fogedby94}
Fogedby HC, Langevin equations for continuous time L\'{e}vy flights. Phys Rev E 1994; 50:1657 

\bibitem{RGarrappa16} 
Garrappa R, Mainardi F, Maione G. Models of dielectric relaxation based on completely monotone functions. Frac Calc Appl Anal 2016; 19:1105 
; corrected version available in {\tt arXiv: 1611.04028}

\bibitem{AGiusti20} Giusti A, Colombaro I, Garra R, Garrappa R, Polito F, Popolizio M, Mainardi F. A practical guide to Prabhakar fractional calculus. Frac Calc Appl Anal 2020; 23:9 

\bibitem{AGiusti20a} Giusti A. General fractional calculus and Prabhakara's theory. Comm Nonlinear Sci Numer Simulat 2020; 83:105114 

\bibitem{Gloeckle} 
Gl\"{o}ckle WG, Nonnenmacher TF. Fox function representation of non-Debye relaxation processes. J Stat Phys 1993; 71:741 

\bibitem{RGorenflo14} Gorenflo R, Kilbas AA, Mainardi F,  Rogosin SV. Mittag-Leffler Functions, Related Topics and Applications. Springer, New York; 2014

\bibitem{KGorska18} 
G\'{o}rska K, Horzela A, Bratek {\L}, Penson KA, Dattoli G. The Havriliak-Negami relaxation and its relatives: the response, relaxation and probability density functions. J Phys A: Theor Math 2018; 51:135202

\bibitem{KGorska19} G\'{o}rska K, Horzela A, Pog\'{a}ny TK. A note on the article "Anomalous relaxation model based on the fractional derivative with a Prabhakar-like kernel" $[$Z. Angew. Math. Phys. $(2019)~ { 70}: 42]$. Z Angew Math Phys 2019; 70:141 

\bibitem{KGorska20} 
G\'{o}rska K, Horzela A. The Volterra type equation related to the  non-Debye relaxation. Comm Nonlinear Sci Numer Simulat 2020; 85:105246 

\bibitem{KGorska20a} 
G\'{o}rska K, Horzela A, Lattanzi A, Pog\'{a}ny TK. On the complete monotonicity of the three parameter generalized Mittag-Leffler function $E_{\alpha, \beta}^{\gamma}(-x)$. Appl Anal Discret Math 2021; 15:118 

\bibitem{KGorska21} 
G\'{o}rska K, Horzela A. Non-Debye Relaxations: Two types of memories and their Stieltjes character, Mathematics 2021; 9:477 

\bibitem{KGorska21a} G\'{o}rska K, Horzela A, Pog\'{a}ny TK. Non-Debye relaxations: smeared time evolution, memory effects, and the Laplace exponents, Comm Nonlinear Sci Numer Simulat  2021; 99:105837 

\bibitem{GGripenberg} 
Grippenberg G, Londen SO, Staffans OJ. Volterra Integral and Functional Equations. Cambridge University Press, Cambridge; 1990

\bibitem{Hanyga1} 
Hanyga A, Seredy\'nska M. On a Mathematical Framework for the Constitutive Equations of Anisotropic Dielectric Relaxation. J Stat Phys 2008; 131:269 

\bibitem{AHanyga20} 
Hanyga A. A comment on a controversial issue: A generalized fractional derivative cannot have a regular kernel. Frac Calc Appl Anal 2020; 23:211 

\bibitem{RHilfer02b} 
Hilfer R. Fitting the excess wing in the dielectric $\alpha$-relaxation of propylene carbonate. J Phys: Condens Matter 2002; 14:2297 

\bibitem{RHilfer02c} 
Hilfer R. Experimental evidence for fractional time evolution in glass forming materials. Chem Phys 2002; 284:399 

\bibitem{RHilfer02} 
Hilfer R. Analytical representations for relaxation functions of glasses. J Non-Cryst Solids 2002; 305:122 

\bibitem{RHilfer02a} 
Hilfer R. $H$-function representations for stretched exponential relaxation and non-Debye susceptibilities in glassy systems. Phys Rev E 2002; 65:061510  

\bibitem{RHilfer17} Hilfer R. Composite continuous time random walks. Eur Phys J B 2017; 90:233 

\bibitem{AJonscher92} 
Jonscher AK. The universal dielectric response and its physical significance. IEEE Transactions on Electrical Insulation 1992; 27:407 

\bibitem{ANKochubei11} 
Kochubei AN. General fractional calculus, evolution equations, and renewal processes. Integr Equ Oper Theory 2011; 71:583 

\bibitem{JDEKonhauser67} Konhauser JDE. Biorthogonal polynomials suggested by the Laguerre polynomials. Pacific J Math 1967; 21:303 

\bibitem{RRNigmatulin16} Nigmatulin RR, Khamzin AA, Baleanu D. On the Laplace integral representation of multivariable Mittag-Leffler functions in anomalous relaxation. Math Meth Appl Sci 2016; 39:2983 

\bibitem{RGarrappa11} 
Mainardi F, Garrappa R. On complete monotonicity of the Prabhakar function and non-Debye relaxation in dielectrics.  J Comp Phys 2015; 293:70 

\bibitem{AMMathai08} Mathai AM, Haubold HJ. Special functions for applied scientists. Springer, New York; 2008 

\bibitem{Ozar} \"Ozarslan MA, K\"urt C. Bivariate Mittag--Leffler functions arising in the solutions of convolution integral equation with $2D$--Laguerre--Konhauser polynomials in the kernel. Appl Math Comput 2019; 347:631 

\bibitem{TRPrabhakar69} Prabhakar TR. A singular integral equation with a generalized Mittag Leffler function in the kernel. Yokohoma Math J 1971; 19:7 

\bibitem{APPrudnikov-v2} Prudnikov AP, Brychkov YuA, Marichev OI. Integrals and Series. Special Functions. Vol. 2. Gordon and Breach, Amsterdam; 1998

\bibitem{TSandev19} Sandev T, Tomovski \v Z. Fractional Equations and Models. Theory and Applications. Springer, New York; 2019

\bibitem{Schilling} 
Schilling RL. An introduction to L\'evy and Feller processes. In: From L\'evy--type processes to parabolic SPDEs. Adv. Courses Math., pp. 1--126. Birkh\"auser -- Springer, Cham (2016)

\bibitem{RSchilling10} 
Schilling RL, Song R, Vondra\v cek  Z. Bernstein Functions. De Gruyter, Berlin; 2010

\bibitem{HMSrivastava82} Srivastava HM. Some biorthogonal polynomials suggested by the Laguerre polynomials. Pacific J Math 1982; 98:235 

\bibitem{AStanislavsky15} 
Stanislavsky A, Weron K, Weron A. Anomalous diffusion approach to non-exponential relaxation in complex physical systems. Comm Nonlinear Sci Numer Simulat 2015; 24:117 

\bibitem{SW16} 
Stanislavsky A, Weron K. Atypical case of the dielectric relaxation responses and its fractional kinetic equation. Frac Calc Appl Math 2016; 19:212 

\bibitem{AStanislavsky17} 
Stanislavsky A, Weron  K. Stochastic tools hidden behind the empirical dielectric relaxation laws. Rep Prog Phys 2017; 80:036001 

\bibitem{AStanislavsky19_1} 
Stanislavsky A, Weron K. Fractional-calculus tools applied to study the nonexponential relaxation in dielectrics. In: Tarasov VE, editor. Handbook of fractional calculus with applications in physics, Part B, Vol. 5. De Gruyter, Berlin; 2019. p. 53--70. 

\bibitem{SW21}
Stanislavsky A, Weron K. Duality in fractional systems, Comm Nonlinear Sci Numer Simulat 2021; 101:105861 

\bibitem{Tomovski} 
Tomovski \v Z, Pog\'any TK, Srivastava HM. Laplace type integral expression for a certain three-parameter family of generalized Mittag-Leffler functions with application involving complete monotonicity. J Franklin Inst 2014; 351:5437 

\bibitem{L1} Details of these distributions are irrelevant, however for applications to the relaxation phenomena it is usually assumed that we deal with $\alpha$-stable distributions; final results come from the generalized limit theorems.

\bibitem{L2} From the probabilistic point of view the Bernstein functions may be identified as subordinators and thus it cannot be strange that they play the crucial role in stochastic analysis of relaxation and anomalous diffusion.

\bibitem{L3} CMFs provide us also with an example of the fading memory concept proposed by L. Boltzmann and reintroduced to physics through applications in rheology and elasticity theory, see e.g. \cite{RSAnderssen02,RSAnderssen02a}.

\bibitem{L4} \label{f3} Throughout the paper the superscript $\,\widehat{}\,$ denotes the Laplace transform:  $\widehat{h}(s) = \mathcal{L}[h(t); s] = \int_{0}^{\infty} \E^{-s t} h(t) \D t$; $h(t)$ is the inverse Laplace transform given as $h(t) = \mathcal{L}^{-1}[\widehat{h}(s); t] = \int_{L} \E^{ s t} \widehat{h}(s) \D s/(2\I\!\pi)$ where $L$ is a Bromwich contour which leaves all singularities of $\widehat{h}(s)$ left to it.

\bibitem{L5} Main properties of the three parameter Mittag--Leffler function $E_{\mu, \nu}^{\gamma}(z)$ are listed in \ref{appA}.

\end{thebibliography}
\end{document}